\documentclass[10pt,doublecolumn,twoside]{IEEEtran}
\IEEEoverridecommandlockouts
\usepackage{cite}
\usepackage{amsmath,amssymb,amsfonts}
\usepackage{algorithmic}
\usepackage{graphicx}
\usepackage{textcomp}
\usepackage{epstopdf}
\usepackage{float}
\usepackage{esint}
\usepackage{subfigure}
\usepackage{color}
\usepackage{mathrsfs}
\usepackage{algorithm}
\usepackage{algorithmic}
\usepackage{bm}
\usepackage{cases}
\usepackage[final]{changes}
\usepackage{setspace}
\usepackage{makecell}
\usepackage{pifont}
\usepackage{threeparttable}

\newtheorem{theorem}{\bf Theorem}

\def\BibTeX{{\rm B\kern-.05em{\sc i\kern-.025em b}\kern-.08em
    T\kern-.1667em\lower.7ex\hbox{E}\kern-.125emX}}

\makeatletter

\newcommand{\Rmnum}[1]{\expandafter\@slowromancap\romannumeral #1@}
\makeatother

\newcommand{\ls}[1]
    {\dimen0=\fontdimen6\the\font
     \lineskip=#1\dimen0
     \advance\lineskip.5\fontdimen5\the\font
     \advance\lineskip-\dimen0
     \lineskiplimit=.9\lineskip
     \baselineskip=\lineskip
     \advance\baselineskip\dimen0
     \normallineskip\lineskip
     \normallineskiplimit\lineskiplimit
     \normalbaselineskip\baselineskip
     \ignorespaces
    }



\begin{document}
\title{Statistical Age of Information: A Risk-Aware Metric and Its Applications in Status Updates}


\author{Yuquan Xiao,~\IEEEmembership{Graduate Student Member,~IEEE}, Qinghe Du,~\IEEEmembership{Member,~IEEE}, and George K. Karagiannidis,~\IEEEmembership{Fellow,~IEEE}
\thanks{\ls{.5}The work reported in this paper was supported in part by the National Key R\&D Program of China under the Grant No. 2020YFB1807700, in part by the National Natural Science Foundation of China under the Grant No. 62071373, in part by the Key Research and Development Program of Shaanxi Province under the Grant No. 2023-GHZD-05, and in part by the Innovation Capability Support Program of Shaanxi under the Grant No. 2021TD-08. The correspondence author of this paper is Qinghe Du (e-mail:duqinghe@mail.xjtu.edu.cn).

Yuquan Xiao and Qinghe Du are with the School of Information and Communications Engineering, Xi'an Jiaotong University, Xi'an 710049, China, and Shaanxi Smart Networks and Ubiquitous Access Research Center, Xi’an 710049, China.

G. K. Karagiannidis is with Department of Electrical and Computer Engineering, Aristotle University of Thessaloniki, Greece and also with Cyber Security Systems and Applied AI Research Center, Lebanese American University (LAU), Lebanon.
}
\vspace{-30pt}
}

\maketitle
\begin{abstract}
Age of information (AoI) is an effective measure to quantify the information freshness in wireless status update systems. It has been further validated that the peak AoI has the potential to capture the core characteristics of the aging process, and thus the average peak AoI is widely used to evaluate the long-term performance of information freshness. However, the average peak AoI is a risk-insensitive metric and therefore may not be well suited for evaluating critical status update services. Motivated by this concern, and following the spirit of entropic value-at-risk (EVaR) in the field of risk analysis, in this paper we present a concept, termed \emph{Statistical AoI}, for providing a unified framework to guarantee various requirements of risk-sensitive status-update services with the demand on the violation probability of the peak age. In particular, as the constraint on the violation probability of the peak age varies from loose to strict, the statistical AoI evolves from the average peak AoI to the maximum peak AoI. We then investigate the statistical AoI minimization problem for status updates over wireless fading channels. It is interesting to note that the corresponding optimal sampling scheme varies from step to constant functions of the channel power gain with the peak age violation probability from one to zero. We also address the maximum statistical AoI minimization problem for multi-status updates with time division multiple access (TDMA), where longer transmission time can improve reliability but may also cause the larger age. By solving this problem, we derive the optimal transmission time allocation scheme. Numerical results show that our proposals can better satisfy the diverse requirements of various risk-sensitive status update services, and demonstrate the great potential of improving information freshness compared to baseline approaches.
\end{abstract}




\section{Introduction}
\IEEEPARstart{I}{International} Telecommunication Union (ITU) recommends the upcoming sixth generation of wireless mobile networks (6G) to advance six significant usage scenarios, including massive communication, hyper-reliable and low-latency communication, ubiquitous connectivity, artificial intelligence and communication, integrated sensing and communication, and immersive communication~\cite{liu2023vision}, some of which refer to time-sensitive applications, and thus enhancing the experience of time-sensitive applications remains as a key objective of future wireless mobile networks.

Wireless status update systems are one of the representative time-sensitive applications where the real-time status of a source node is desired to be known at the destination node in a timely manner, and this process is going on everywhere in our lives tightly coupled with a digital world, such as speed monitoring in autonomous driving, location sharing among friends with social applications, actuator control in industrial automation, situational sensing for emergency rescue, etc.~\cite{Cao2023DT,Yates2021Age,Yao2022The}. In particular, some status update systems are risk-sensitive, where the critical status information on the target side should always be as fresh as possible. Otherwise, even a small amount of outdated information could cause undesirable system behavior and even damage.

How to evaluate the freshness of status information is extremely important in wireless status update systems. Age of Information~(AoI) is a recently emerging and attractive concept in this area, which is defined as how much time has elapsed since the last received status information was generated at the source~\cite{Kaul2012Real}. Following this idea, AoI is a sawtooth function with respect to time evolution. Unlike traditional delay metrics~\cite{Cheng2022Adaptive,Zhao2023Improved,Zhu2023Joint,Xiong2018Cross}, AoI is determined not only by the end-to-end delay over the network, but also by the sampling rate at the source. For example, low sampling rate may be advantageous for low end-to-end delay due to low traffic load. But in the meantime, a low sampling rate may still lead to the decision delay even with the stale information at the destination. By fine-tuning the sampling rate at the source, the information freshness can be significantly enhanced, with the associated cost being comparatively modest in contrast to implementing semantic communications for load reduction aimed at improving information freshness~\cite{Fu2023Scalable}.

AoI has attracted a lot of research attention, where the average AoI has been widely used to evaluate the long-term performance of information freshness in various status update systems~\cite{Zheng2019Closed,Ge2024AoI,Bao2022Average,Lin2021Cooperative}. However, some researches point out that the derivation of the average AoI is often very complicated due to the coupling between the inter-arrival time and the system time of status data packets~\cite{Yates2021Age,Yang2021Understanding}. In contrast, since AoI is a sawtooth function of time, the peak AoI is able to capture the core characteristics of the aging process~\cite{Costa2014Age}. Then, the average peak AoI can serve as an alternative long-term performance metric instead of the average AoI, which avoids the complicated derivation caused by the correlations between the inter-arrival time and the system time. Although the average statistic of AoI can well reflect the long-term performance of information freshness and is often used as an objective function to guide resource allocation in wireless status update systems, it cannot meet the requirements of critical status update systems. This is because even if the average AoI or the average peak AoI is minimized, the instantaneous age or the peak age often fluctuates due to the time-varying environments such as the fading of wireless channels, and thus we do not know what percentage of the peak age is beyond the average AoI or the average peak AoI. Of course, there remains the uncontrollable risk that the information age is often large, causing the unintended consequences, for example, the delay of image rendering in the metaverse applications will lead to poor quality of user experience~\cite{xiao2023metaverse}. Some literature also considers minimizing the maximum peak AoI to mitigate the worst-case influence of outdated information~\cite{Chen2023Peak,He2016On,Cao2020Peak}. However, minimizing the maximum peak AoI often creates a bias that allocates too many resources to the worst case, and overall performance is significantly degraded. In summary, average AoI or average peak AoI is risk insensitive, while maximum peak AoI is oversensitive for certain critical status update systems.

To meet the requirements of various critical status update systems, the minimum achievable peak AoI with a given violation probability can be used as an alternative, yet more comprehensive metric, where the level of violation probability can be linked to the risk sensitivity of the status update systems. In particular, the higher the risk sensitivity, the lower the violation probability should be, and vice versa. Furthermore, this metric is expected to be extended to reflect the statistical characteristics of the peak AoI beyond a certain threshold, e.g., the average of the peak AoI at the tail, which can better indicate the overall performance as well as the risk level, i.e., the given violation probability. However, since the probability distribution function (PDF) of the peak AoI often lacks a closed-form expression, it is difficult to derive the minimum achievable peak AoI with a given violation probability, and even harder to improve it. Research in risk analysis~\cite{pritsker1997evaluating,rockafellar2000optimization,ahmadi2012entropic} provides us with an effective way to tackle this problem. Specifically, the above minimum achievable peak AoI can be seen as an application of Value at Risk (VaR) in the field of risk analysis~\cite{pritsker1997evaluating}. However, the use of VaR cannot reflect the extent of peak AoI that may be experienced beyond the minimum achievable peak AoI. In this sense, the concept of conditional value-at-risk (CVaR) in risk analysis can be used, which is defined as the average of the risk value at the tail~\cite{rockafellar2000optimization}. However, a recent study shows that the derivation of CVaR is also computationally expensive. To overcome this problem, entropic value-at-risk (EVaR) has been proposed in~\cite{ahmadi2012entropic} by exploiting the Chernoff-Cram\'{e}r bound, which is an upper bound of CVaR and is efficiently computable. Motivated by EVaR, in this paper we introduce the concept of statistical AoI, which corresponds to a tight upper bound of the average peak AoI at the tail distribution, with the promise of satisfying the specified risk-sensitive application's requirement. Given the mathematical properties of EVaR, the statistical AoI is more convenient to use, especially for theoretical analysis, than the minimum achievable peak AoI. In addition, it is worth mentioning that as the probability of violation approaches one and zero, the statistical AoI reduces to the average peak AoI and the maximum peak AoI, respectively.

\subsection{Related Work}
We review recent literatures on long-term AoI metrics, as well as corresponding optimization approaches and applications for various wireless status update systems. The seminal work of AoI~\cite{Kaul2012Real} presented the definition of the average AoI and proposed to use graphical methods to compute it, where the long-term AoI performances of some typical queuing systems such as M/M/1 and M/D/1 with first-in-first-out discipline have been studied accordingly. In M/M/1 systems where sometimes the renewal of information is missed, the authors of \cite{Lin2022Age} evaluated the corresponding influence by deriving an upper bound on average AoI. In addition to using the average AoI to evaluate AoI performance, much research effort has been devoted to minimizing it in order to manage and allocate system resources. The authors of~\cite{zhang2023AoI} studied the average AoI minimization problems over wireless energy-constrained sensor networks and derived the corresponding sampling and power allocation schemes. In~\cite{Fountoulakis2023Scheduling}, the authors studied how to satisfy the quality of service (QoS) requirements of mixed traffic, including AoI-oriented service and throughput-demanding service, by minimizing the average AoI under throughput constraint. The literature~\cite{Huang2022Age} discussed the minimization of the weighted linear combination of the average AoI and the total energy consumption for status updates over wireless fading channels, where retransmission was considered and the optimal power control policy was obtained. The authors of~\cite{Hatami2021AoI} studied the average AoI minimization problems in cache-assisted energy harvesting networks, which is modeled as a Markov decision process, and the reinforcement learning algorithm was used to identify the solutions.

As mentioned above, sometimes the deviation of the average AoI is very challenging. In this regard, the authors of~\cite{Costa2014Age} introduced the concept of average peak AoI instead of average AoI to evaluate the long-term freshness performance of status update systems. Reference~\cite{Zhao2022Information} revealed that the optimal channel access strategy is the same when using either average AoI or average peak AoI as the optimization objective in Poisson networks.
The authors of~\cite{Cao2023Multiplexing} analyzed the achievable peak AoI in dual-link status updates over wireless fading channels and showed that diversity gain is more desirable for the low signal-to-noise ratio (SNR) case, while multiplexing gain is more desirable for the high SNR scenario. The work~\cite{Champati2021Minimum} used the average peak AoI to evaluate the information freshness performance of single-source-single-service systems for general service-time distributions with consideration of service preemptions. The results showed that a higher gain of average peak AoI can be achieved when the tail of a service-time distribution is heavier. The authors of~\cite{Jia2023Service} investigated the average peak AoI minimization problem and devised the optimal service rate allocation scheme in a dual-sensor monitor system, where two state processes are jointly updated by two sensors.

Besides efforts to study the average statistics of AoI, some researches also considered using the maximum peak AoI to quantify the information freshness performance. Reference~\cite{Chen2023Peak} investigated the maximum-peak AoI minimization problems in wireless-powered edge networks, where an efficient charging and data transmission scheme was developed. The maximum peak AoI was targeted to be minimized for the multi-user multi-link scheduling scenarios, which was NP-hard, but solved by using mixed integer linear programming approaches~\cite{He2016On}.
The authors of~\cite{Cao2020Peak} proposed the optimal flight trajectory as well as the transmission power scheme to reduce the maximum peak AoI in unmanned aerial vehicle (UAV)-assisted relay transmissions. There are also other studies on optimizing the tail of the AoI distribution in certain specified networks. Reference~\cite{Wang2023Statistical} presented an upper-bound of the peak age violation probability by using the large deviations principle. The authors of~\cite{Champati2021Statistical} investigated the problem of minimizing the tail of the AoI distribution in multi-hop status update scenarios with periodic packet arrivals, where the derivation of the exact expression of the AoI distribution proved to be difficult. The authors of~\cite{Zhou2020Risk} used the conditional value-at-risk (CVaR) measure for AoI and then used it to design the optimal update scheme. In~\cite{de2023Risk}, risk and cost analyses for risk-sensitive state updates with random arrivals were formulated as a Markov decision process (MDP) problem, where the frequency of visiting risky states was minimized using the reinforcement learning approach. We find that there is little work done on risk-aware status updates and several related studies are conducted in specified scenarios. Therefore, it is essential to build a unified framework to guide the design and evaluate the information freshness performance for various risk-aware status update systems.

\subsection{ Main Contributions}
\label{sec:contribution}
In this paper, we focus on how to satisfy the diverse requirements of different risk-sensitive status update systems and provide corresponding design methodologies. The main contributions of our work are summarized below:

\begin{itemize}
	\item The concept of \emph{statistical AoI} is introduced. Following the definition of EVaR in the field of risk analysis, statistical AoI is the tight upper bound derived from the Chernoff-Cram\'{e}r bound on the average peak age of the tail distribution, with the probability of violation of the peak age as a predetermined parameter in the definition. This concept has the potential to characterize various risk-sensitive requirements for status update systems by restricting the value of the violation probability to the given peak age. It is worth noting that as the violation probability varies from one to zero, the statistical AoI metric reduces to the average peak AoI and the maximum peak AoI, respectively. Accordingly, since the Chernoff-Cram\'{e}r bound is used, a key parameter called the AoI exponent is introduced, which can indicate the exponentially varying tendency of the violation probability against the peak age. In general, the smaller the violation probability at peak age, the larger the AoI exponent.
	\item To address the important role that the statistical AoI metric can play, we formulate and study the statistical AoI minimization problem over wireless fading channels, where the independent optimization variables include the AoI exponent and the sampling rate as functions of the channel power gain. We propose a two-step method to solve the problem. In the first step, we consider the AoI exponent as a constant and the problem reduces to a fractional programming problem with respect to the sampling rate. By applying Dinkelbach's transform to this reduced problem, the optimal sampling scheme can be effectively derived. We then search for the optimal AoI exponent subject to the corresponding optimal sampling scheme derived in the first step by bisection. Insightful results are obtained that as the violation probability varies from one to zero, the optimal sampling scheme changes from a step function of the channel power gain to a constant function.
	\item We also study the maximum-statistical-AoI minimization problem for multi-source status updates with time division multiple access (TDMA), where the transmission time allocated to each source can be adjusted. It is worth noting that the longer transmission time increases the reliability and may also cause the large information age. Since the risk requirement of each source is different from others, we specifically study how to allocate transmission time for each source to minimize the maximum statistical AoI. The formulated problem is challenging to solve simply. Alternatively, we derive an efficient algorithm that first considers the single-source case with the constrained maximum transmission time, and then applies the bisection method to find the optimal transmission time for the multi-source case.
	\item Numerical results show that our proposals for status updates over wireless fading channels can achieve the lower statistical AoI compared with existing average/maximum-peak AoI-oriented schemes, and thus better satisfy the requirements of various risk-aware status update systems. Moreover, for multi-status updates, our proposals can more efficiently utilize limited resources to satisfy users' risk-sensitive requirements while improving the overall performance in terms of information freshness.
\end{itemize}

The rest of this paper is organized as follows. Section~\ref{sec:statistical_aoi} presents the definition of statistical AoI. The statistical AoI minimization problem over the wireless fading channels is formulated and solved in Section~\ref{sec:statistical_aoi_phy}. The maximum-statistical-AoI minimization problem for multi-status updates is studied in Section~\ref{sec:statistical_aoi_mac}. Section~\ref{sec:numerical_results} presents the performance of our proposals in comparison with existing schemes. Finally, the highlights of this paper are concluded with Section~\ref{sec:conclusions}.

\section{The Concept of Statistical AoI for Monitoring Systems}
\label{sec:statistical_aoi}
\begin{figure}
\centering
\includegraphics[scale = 0.43]{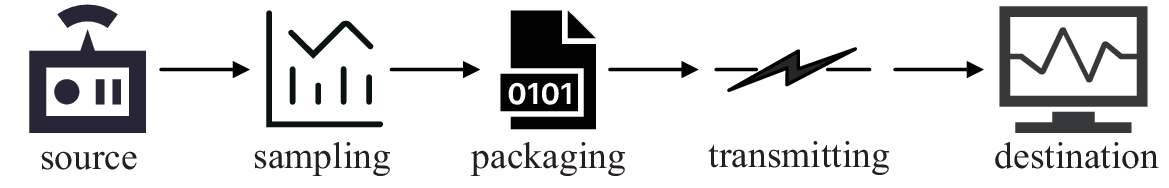}
\caption{A typical status updating process in monitoring systems.}
\label{fig:monitoring}
\end{figure}
\begin{figure*}
\centering
\includegraphics[scale = 1.1]{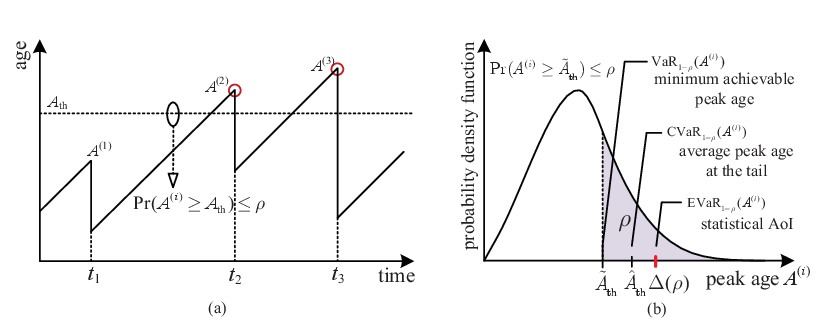}
\caption{(a) The AoI process at the side of destination, where $A^{(i)}$ represents the $i$th peak age and $A_{\rm th}$ denotes the lower bound of peak age satisfying ${\rm Pr}(A \geq A_{\rm th}) \leq \rho$, and (b) the relationships among VaR, CVaR, and EVaR in terms of peak age.}
\label{fig:statistical_aoi}
\end{figure*}
As shown in Fig.~\ref{fig:monitoring}, a status monitoring system includes a source that monitors a particular status of interest of an object and a destination that stands by to receive the monitored status sent by the source via wireless transmissions. The source follows a predetermined procedure to sample the real-time status of interest, package the sample data, and deliver the corresponding data packet to the destination. Due to various uncertain factors, such as medium access contention among multiple transmitters, fading of wireless channels, interference, transmission errors, etc., the status information successfully received at the destination can easily become stale. Age of Information (AoI) can be used to quantify the freshness of information at the destination side, which is defined as the amount of time elapsed since the last received information was originally generated. Following this definition, we can illustrate the age process in Fig.~\ref{fig:statistical_aoi}(a), where the age at the destination side increases linearly within the interval between the arrival of two adjacent status packets. The age at the destination drops sharply, resulting in a sawtooth edge shown in Fig.~\ref{fig:statistical_aoi}(a), each time a new status packet is correctly received. The age is then immediately updated to the current time minus the generation time of this newly arrived status packet. Existing research has shown that the peak age can be used to capture the main mathematical properties of the time-varying age process~\cite{Costa2014Age}. Therefore, in this paper, we focus on the corresponding stochastic sequence of peak ages of an age process, i.e., $\{A^{(1)}, A^{(2)}, A^{(3)}, ...\}$ in Fig.~\ref{fig:statistical_aoi}(a), where $A^{(i)}$ denotes the $i$th age peak. In existing research~\cite{Costa2014Age}, the average peak age, defined as $\lim_{N\rightarrow\infty}\frac{1}{N}\sum_{i = 1}^{N}A^{(i)}$, is often used as a metric to evaluate the ability of a status update system to ensure information timeliness. However, for some critical status updates, such as the speed of neighboring vehicles in autonomous driving, the average peak AoI is insufficient to reflect the satisfaction level of users. In particular, the average peak AoI cannot indicate the percentages of peak ages beyond the average peak AoI level, where even not too many large peak ages can have destructive consequences. Therefore, a risk-aware long term metric should be developed to optimize critical state update systems. To this end, the minimum achievable peak AoI, denoted by $\widetilde A_{\rm th}$, with a given violation probability, denoted by $\rho$, can be considered as a metric to satisfy the requirements of risk-sensitive status update applications, which is defined as follows:

\begin{align}\label{var_aoi}
\widetilde A_{\rm th} = \min\left\{A_{\rm th} | {\rm Pr}\left(A^{(i)} \geq A_{\rm th}\right) \leq \rho\right\},
\end{align}
where $\min\{\cdot\}$ represents the minimum value in a set and ${\rm Pr}(\cdot)$ denotes the probability of an event. However, using $\widetilde A_{\rm th}$ as the metric cannot effectively reflect the tail-distribution features of the peak age. We expect that the peak age at the tail is also concentrated around a small value. To achieve this goal, the average peak age at the tail, denoted by $\widehat A_{\rm th}$, can be considered as a powerful risk-aware metric, that is,
\begin{align}
\widehat A_{\rm th} = \min_{A_{\rm th}} \left\{A_{\rm th} + \frac{1}{\rho}\mathbb{E}\left[\max\left(A^{(i)} - A_{\rm th}, 0\right)\right]\right\},
\end{align}
where $\mathbb{E}[\cdot]$ denotes the mathematical expectation and $\max(\cdot, \cdot)$ indicates the maximum value between two values. It is worth mentioning that, as illustrated in Fig.~\ref{fig:statistical_aoi}(b), the minimum achievable peak AoI $\widetilde A_{\rm th}$ and the average peak age at the tail $\widehat A_{\rm th}$ correspond to the concept of value-at-risk (VaR) and conditional value-at-risk (CVaR) in the field of risk analysis~\cite{pritsker1997evaluating,rockafellar2000optimization}, respectively. However, the study~\cite{ahmadi2012entropic} shows that computing CVaR is often difficult and inefficient. To tackle this issue, entropic value-at-risk (EVaR) is proposed by using the Chernoff-Cram\'{e}r bound, which is an upper-bound of CVaR and can be efficiently computed~\cite{ahmadi2012entropic}.
Following the definition of EVaR in the field of risk analysis, we introduce the concept of statistical AoI, which is a tight upper-bound of the average peak age at the tail distribution. We represent the statistical AoI with respect to $\rho$ as $\Delta(\rho)$. Mathematically, $\Delta(\rho)$ is defined as
\begin{align}\label{eq:definition_statistical_aoi}
\Delta(\rho) = \min_\theta \frac{1}{\theta}\log\left\{\frac{1}{\rho}M_A(\theta)\right\},
\end{align}
where $\theta$ is called AoI exponent and $M_A(\theta)$ is moment generating function (MGF) of the peak age, i.e., the mathematical expectation of $e^{\theta A}$. Here, the superscript of $A^{(i)}$ is omitted. Then, the probability of the peak age exceeding the statistical AoI is accordingly given as follows:
\begin{align}\label{eq:probability}
{\rm Pr}\left(A \geq \Delta(\rho)\right) \leq \rho.
\end{align}
\begin{table*}
    \centering
    \footnotesize
    \caption{Long-Term Performance Metrics in Terms of AoI}\label{tab:aoi_metric}
    \begin{threeparttable}
    \begin{tabular}{|c|c|c|}
    \hline
    \textbf{Metric} &\textbf{Definition} & \textbf{Features}\\ \hline\hline
    Average AoI & $\lim_{\mathcal{T}\rightarrow\infty}\frac{1}{\mathcal{T}}\int_0^\mathcal{T} a(t)dt$ & \footnotesize\makecell[l]{\emph{Pros:} 1) Well evaluate the overall AoI performance\\\emph{Cons:}\! 1) Sometimes hard to compute; 2) Risk-insensitive}\\ \hline
    Average peak AoI & $\lim_{N\rightarrow\infty}\frac{1}{N}\sum_{i = 1}^{N}A^{(i)}$ & \footnotesize\makecell[l]{\emph{Pros:} 1) Well evaluate the overall AoI performance\\\emph{Cons:}\! 1) Risk-insensitive}\\ \hline
    Maximum peak AoI & $\max\left\{A^{(1)},A^{(2)},A^{(3)},...\right\}$ &\makecell[l]{\emph{Pros:} 1) Take the worst case into consideration\\\emph{Cons:}\! 1) Risk-oversensitive}\\ \hline
    \makecell{Minimum peak AoI\\with tolerable violation\\probability (VaR)} & $\min\left\{A_{\rm th} | {\rm Pr}\left(A^{(i)} \geq A_{\rm th}\right) \leq \rho\right\}$ & \footnotesize\makecell[l]{\emph{Pros:} 1) Risk-sensitive\\\emph{Cons:}\! 1) Overlook the statistics of the peak age at the tail; \\2) Hard to compute}\\ \hline
    \makecell{Average peak AoI\\at the tail (CVaR)} & $\min_{A_{\rm th}} \left\{A_{\rm th} + \frac{1}{\rho}\mathbb{E}\left[\max(A^{(i)} - A_{\rm th}, 0)\right]\right\}$ & \footnotesize\makecell[l]{\emph{Pros:} 1) Risk-sensitive; 2) Take into account the statistics\\of the peak age at the tail\\\emph{Cons:}\! 1) Hard to compute}\\ \hline
     Statistical AoI (EVaR) & $\min_\theta \frac{1}{\theta}\log\left\{\frac{1}{\rho}M_A(\theta)\right\}$ & \footnotesize\makecell[l]{\emph{Pros:} 1) Risk-sensitive; 2) Take into account the statistics\\of the peak age at the tail; 3) Effectively computable}\\ \hline
    \end{tabular}
    \begin{tablenotes}
    \item[*] $a(t)$ denotes the age process, $A^{(i)}$ denotes the $i$th age peak, and $\rho$ denotes the violation probability of the peak age.
    \end{tablenotes}
    \end{threeparttable}
\end{table*}
Owing to the differentiable properties of MGF and the explicit mappings among $A$, $\Delta(\rho)$, and $\rho$, the statistical AoI potentially eliminates the disadvantages of $\widetilde A_{\rm th}$ as well as $\widehat A_{\rm th}$ and is particularly suited to serve as a risk-aware metric or objective to guide the designing of status update systems. Moreover, it can be shown that the average peak age and the maximum peak age are the special cases of statistical AoI, where we have
\begin{align}
\lim_{\rho\rightarrow 1} \Delta(\rho) = \mathbb{E}[A]\mbox{~and~}\lim_{\rho\rightarrow 0} \Delta(\rho) = \max(A),
\end{align}
respectively. In summary, we conclude the main long-term AoI metrics with their definitions and features in Table~\ref{tab:aoi_metric}.

\section{Application of Statistical AoI at Physical Layer: Critical Status Update over Wireless Fading Channels}
\label{sec:statistical_aoi_phy}
We in this section present how the concept of statistical AoI can be used to guide the designs of sampling scheme for critical status update applications over wireless fading channels.
\subsection{System Description}
\label{sec:system_model_fading}
Consider a status update system where the status packets are delivered from the sensing source to the destination over a wireless fading channel. Specifically, the channel coherence time and the channel power gain are denoted by $T$ and $\gamma$, respectively. We assume that the channel follows the block fading model, i.e., the channel power gain remains constant within a coherence time and varies independently over different channel coherence times. For the generation of status packets in the source, the generate-at-will model is adopted, where the packets can be generated immediately as soon as we have a will~\cite{Bhat2021Throughput}. We denote by $\lambda(\gamma,\rho)$ the sampling rate with respect to the channel power gain $\gamma$, which is assumed to be known by the source, and the given violation probability $\rho$. Each sampling step results in one status packet. The transmission time for a status packet is represented as $\tau$, which is less than $T$. Then, based on the Shannon capacity, we must have
\begin{align}
\tau B\log_2(1 + P(\gamma,\rho)\gamma) \geq D,
\end{align}
where the noise power is assumed to be one, $B$ is the bandwidth, $P(\gamma,\rho)$ represents the transmit power varying with $\gamma$ and $\rho$, and $D$ denotes the number of bits of one status packet.

In order to prevent the packet backlog from happening and ensure each status packet timely transmitted once it is generated, We demand that the sampling interval should be larger than transmission time, i.e.,
\begin{align}
\tau\lambda(\gamma,\rho) \leq 1.
\end{align}
Moreover, we require sampling at least once over one channel coherence time, that is,
\begin{align}
T\lambda(\gamma,\rho) \geq 1,
\end{align}
which avoids that the sampling rate is too low such that the status information at the destination is severely outdated. In addition, the average transmit power constraint is considered, i.e.,
\begin{align}\label{eq:average_power_constraint}
\mathbb{E}\left[\tau\lambda(\gamma,\rho)P(\gamma,\rho)\right] \leq \bar{P},
\end{align}
where $\bar{P}$ is the average transmit power of the source. There is a coefficient, i.e., $\tau\lambda(\gamma,\rho)$, at the front of $P(\gamma,\rho)$ in Eq.~(\ref{eq:average_power_constraint}), because only part of time $\tau\lambda(\gamma,\rho)$ is consumed for transmission.

\subsection{Age Processes at the Sides of Source and Destination}
\begin{figure}
\centering
\includegraphics[scale = 0.9]{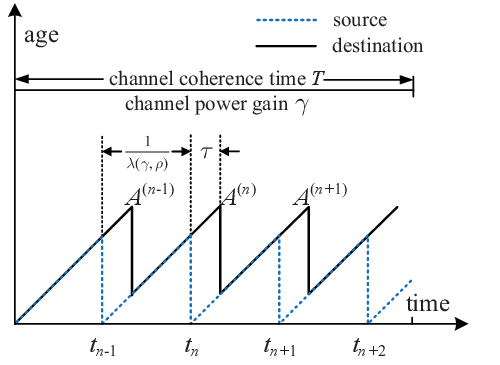}
\caption{AoI processes at the sides of source and destination.}
\label{fig:statistical_aoi_phy}
\end{figure}

The age processes at the sides of source and destination are illustrated in Fig.~\ref{fig:statistical_aoi_phy}. We assume that the duration of a channel coherence time is sufficiently long, which is consistent with typical practical systems, so that multiple sampling steps are performed within a channel coherence interval. As shown in the figure, we denote the time of the $n$th sampling at the source by $t_n$. According to the definition of AoI, the age at the source is set to zero at each sampling instant and then increases with time between two adjacent sampling steps. Since the transmission time is $\tau$, the age at the destination side is set to $\tau$ as soon as a new status packet is received, and then increases between two adjacent status packet arrivals. Then the peak age, $A$, at the destination can be represented as follows:
\begin{align}\label{eq:peak_age_fading}
A = \frac{1}{\lambda(\gamma,\rho)} + \tau.
\end{align}
Since the duration of one channel coherence time is $T$ and the sampling interval is $1/\lambda(\gamma,\rho)$, the amount of age peaks within a time duration $[t,t+T)$ is roughly equal to $T\lambda(\gamma,\rho)$. Additionally, the probability of channel power gain equal to $\gamma$ is $f_\gamma(\gamma)d\gamma$, where $f_\gamma(\gamma)$ is the probability density function (pdf) of the channel power gain. Therefore, the probability of the peak age equal to $(\tau + 1/\lambda(\gamma,\rho))$ can be given by
\begin{align}
{\rm Pr}\left(A = \tau + \frac{1}{\lambda(\gamma,\rho)}\right) = \frac{T\lambda(\gamma,\rho)f_\gamma(\gamma)d\gamma}{\int T\lambda(\gamma,\rho)f_\gamma(\gamma)d\gamma},
\end{align}
Eliminating $T$ from the numerator and the denominator, we obtain
\begin{align}\label{eq:probability_peak_aoi}
{\rm Pr}\left(A = \frac{1}{\lambda(\gamma,\rho)}+ \tau\right) = \frac{\lambda(\gamma,\rho)f_\gamma(\gamma)d\gamma}{\int\lambda(\gamma,\rho)f_\gamma(\gamma)d\gamma}.
\end{align}

\subsection{Statistical AoI Minimization Problem}
According to Eqs.~(\ref{eq:peak_age_fading}) and (\ref{eq:probability_peak_aoi}), the MGF of $A$ can be derived as follows:
\begin{align}\nonumber
\mathbb{E}\left[e^{\theta A}\right]
&= \frac{\int e^{\theta(\tau + \frac{1}{\lambda(\gamma,\rho)})}\lambda(\gamma, \rho)f_\gamma(\gamma)d\gamma}{\int\lambda(\gamma,\rho)f_\gamma(\gamma)d\gamma}\\
&=\frac{e^{\theta\tau}\mathbb{E}[e^{\frac{\theta}{\lambda(\gamma,\rho)}}\lambda(\gamma, \rho)]}{\mathbb{E}[\lambda(\gamma,\rho)]}.
\end{align}
Then, following the definition of statistical AoI in Eq.~(\ref{eq:definition_statistical_aoi}), the statistical AoI of $A$ can be expressed by
\begin{align}\label{eq:statistical_phy}
\Delta(\rho) = \min_\theta \frac{1}{\theta}\log\frac{e^{\theta\tau}\mathbb{E}\left[e^{\frac{\theta}{\lambda(\gamma,\rho)}}\lambda(\gamma, \rho)\right]}{\rho\mathbb{E}[\lambda(\gamma,\rho)]}.
\end{align}
Furthermore, the statistical AoI minimization problem, denoted by \textbf{P0}, for critical status update over wireless fading channels is formulated as follows:
\begin{subequations}
\begin{align}
\textbf{P0}:
\min_{\lambda(\gamma,\rho),P(\gamma,\rho)} \,\,\, &\Delta(\rho),\\
{\rm s.t.}\,\,\,
& \tau B\log_2(1 + P(\gamma,\rho)\gamma) \geq D,\label{eq:constraint_1} \\
& \mathbb{E}\left[\tau\lambda(\gamma,\rho)P(\gamma,\rho)\right] \leq \bar{P},\label{eq:constraint_2} \\
& \tau \lambda(\gamma,\rho) \leq 1,\label{eq:constraint_3} \\
& T\lambda(\gamma,\rho) \geq 1,\label{eq:constraint_4}
\end{align}
\end{subequations}
where each constraint has been described in Section~\ref{sec:system_model_fading}. Our task is to find out the sampling and power allocation solutions to \textbf{P0}, i.e., the schemes to attain the minimum statistical AoI with the given violation probability of the peak age.

\vspace{-10pt}
\subsection{Optimal Sampling}
As the statistical AoI of $A$ specified in Eq.~(\ref{eq:statistical_phy}) is decreasing with respect to the sampling rate $\lambda(\gamma,\rho)$, the feasible solution with the larger sampling rate constrained by Eqs.~(\ref{eq:constraint_1}) to (\ref{eq:constraint_4}) is likely to be closer to the optimum. In order to obtain a larger sampling rate, the corresponding transmit power in (\ref{eq:constraint_2}) should be smaller. Toward this end, the equaling sign ``='' in (\ref{eq:constraint_1}) shall hold for the optimal solution and then (\ref{eq:constraint_1}) can reduce to
\begin{align}\label{eq:channel_inverse}
P(\gamma,\rho) = \frac{1}{\gamma}\left(e^{\bar{D}_\tau}-1\right),
\end{align}
where $\bar{D}_\tau$ is accordingly equal to $D\log 2/(\tau B)$. This indicates that the optimal transmit power follows the channel inverse scheme that the transmit power is proportional to the inverse of the channel power gain. Then, by substituting (\ref{eq:channel_inverse}) into (\ref{eq:constraint_2}), the average transmit power constraint can be rewritten as follows:
\begin{align}
\mathbb{E}\left[\frac{\lambda(\gamma,\rho)}{\gamma}\right]\leq\bar{P}_\tau,
\end{align}
where $\bar{P}_\tau = \bar{P}/(\tau(e^{\bar{D}_\tau}-1))$.
Merging constraints (\ref{eq:constraint_1}) and (\ref{eq:constraint_2}), we equivalently convert \textbf{P0} to \textbf{P1} as follows:
\begin{subequations}
\begin{align}
\textbf{P1}:
\min_{\lambda(\gamma,\rho),\theta} \,\,\, &\frac{1}{\theta}\log\frac{e^{\theta\tau}\mathbb{E}\left[e^{\frac{\theta}{\lambda(\gamma,\rho)}}\lambda(\gamma, \rho)\right]}{\rho\mathbb{E}[\lambda(\gamma,\rho)]},\\
{\rm s.t.}\,\,\,
& \mathbb{E}\left[\frac{\lambda(\gamma,\rho)}{\gamma}\right]\leq\bar{P}_\tau,\\
& \tau\lambda(\gamma,\rho) \leq 1,\\
& T\lambda(\gamma,\rho) \geq 1.
\end{align}
\end{subequations}
For description brevity, we represent the objective of \textbf{P1} by $f(\theta,\lambda(\gamma,\rho))$, which is the sum of two parts, i.e., $1/\theta\log M_A(\theta)$ and $1/\theta\log(1/\rho)$, as follows:
\begin{align}\label{eq:h_function}
f(\theta,\lambda(\gamma,\rho))=\underbrace{\frac{1}{\theta}\log\frac{\mathbb{E}[e^{\frac{\theta}{\lambda(\gamma,\rho)}}\lambda(\gamma,\rho)]}{\mathbb{E}[\lambda(\gamma,\rho)]} + \tau}_{\frac{1}{\theta}\log M_A(\theta)} + \frac{1}{\theta}\log\frac{1}{\rho}.
\end{align}
To jointly optimize the sampling rate and the AoI exponent is very challenging. We propose a two-step method to solve \textbf{P1}. At the first step, we regard the AoI exponent as a constant and find the optimal sampling scheme. Then, we search for the optimal AoI exponent with its corresponding optimal sampling scheme obtained at the first step. Following this procedure, we can write the subproblem of the first step as \textbf{P2} given below:
\begin{subequations}
\begin{align}
\textbf{P2}:
\min_{\lambda(\gamma,\rho)} \,\,\, &\frac{\mathbb{E}[e^{\frac{\theta}{\lambda(\gamma,\rho)}}\lambda(\gamma,\rho)]}{\mathbb{E}[\lambda(\gamma,\rho)]},\\
{\rm s.t.}\,\,\,
& \mathbb{E}\left[\frac{\lambda(\gamma,\rho)}{\gamma}\right]\leq\bar{P}_\tau,\\
& \tau\lambda(\gamma,\rho) \leq 1,\\
& T\lambda(\gamma,\rho) \geq 1,
\end{align}
\end{subequations}
where the objective of \textbf{P2} is the first part of $1/\theta\log M_A(\theta)$ in $f(\theta,\lambda(\gamma,\rho))$ with omitting the operation $1/\theta\log(\cdot)$ because $\theta$ is a constant.

\textbf{P2} is a standard convex-concave factional programming problem, where the objective follows the convex-concave style and the constraints are convex. The Dinkelbach's transforming method~\cite{dinkelbach1967nonlinear} is used to solve this problem. Using the Dinkelbach's transforming, we convert \textbf{P2} to \textbf{P3} as
\begin{subequations}
\begin{align}
\textbf{P3}:
\min_{\lambda(\gamma,\rho)} \,\,\, &\mathbb{E}\left[\lambda(\gamma,\rho)\left(\beta e^{\frac{\theta}{\lambda(\gamma,\rho)}} - 1\right)\right],\\
{\rm s.t.}\,\,\,
& \mathbb{E}\left[\frac{\lambda(\gamma,\rho)}{\gamma}\right]\leq\bar{P}_\tau,\label{eq:c_p3_1}\\
& \tau\lambda(\gamma,\rho) \leq 1,\label{eq:c_p3_2}\\
& T\lambda(\gamma,\rho) \geq 1,\label{eq:c_p3_3}
\end{align}
\end{subequations}
where $\beta$ is a new introduced auxiliary variable. This variable is iteratively updated by
\begin{align}\label{eq:beta}
\beta[i+1] = \frac{\mathbb{E}[\lambda(\gamma,\rho)[i]]}{\mathbb{E}\left[e^{\frac{\theta}{\lambda(\gamma,\rho)[i]}}\lambda(\gamma,\rho)[i]\right]},
\end{align}
where $i$ denotes the $i$th iteration. By solving \textbf{P3} with updated $\beta$ iteratively until the value of $\beta$ converges, the optimal solution of \textbf{P3} is also optimal to \textbf{P2}.

Since \textbf{P2} is a standard convex-concave factional programming problem, \textbf{P3} can be verified to be a convex problem. Thus, we can use the Lagrangian multiplier method to solve it. The Lagrangian function of \textbf{P3}, denoted by $\mathcal{L}(\lambda(\gamma,\rho);\eta,\eta_\tau,\eta_T)$, is constructed as follows:
\begin{align}
\mathcal{L}(\lambda(\gamma,\rho);\eta,\eta_\tau,\eta_T) = \mathbb{E}\left[\lambda(\gamma,\rho)\left(\beta e^{\frac{\theta}{\lambda(\gamma,\rho)}} - 1\right)\right] \nonumber&\\
+\eta\left(\mathbb{E}\left[\frac{\lambda(\gamma,\rho)}{\gamma}\right]- \bar{P}_\tau\right)
+\eta_\tau(\tau\lambda(\gamma,\rho) - 1) \nonumber&\\
+ \eta_T(1 - T\lambda(\gamma,\rho)),&
\end{align}
where $\eta$, $\eta_\tau$, and $\eta_T$ are  multipliers associated with constraints (\ref{eq:c_p3_1}), (\ref{eq:c_p3_2}), and (\ref{eq:c_p3_3}), respectively. Taking the first-order derivative of $\mathcal{L}(\lambda(\gamma,\rho);\eta,\eta_\tau,\eta_T)$ with respect to $\lambda(\gamma,\rho)$, we have
\begin{multline}
\frac{\partial\mathcal{L}(\lambda(\gamma,\rho);\eta,\eta_\tau,\eta_T)}{\partial\lambda(\gamma,\rho)}=\\
\beta e^{\frac{\theta}{\lambda(\gamma,\rho)}}\left(1 - \frac{\theta}{\lambda(\gamma,\rho)}\right) - 1 + \frac{\eta}{\gamma} + \eta_\tau\tau -\eta_TT.
\end{multline}
Then, the Karush-Kuhn-Tucker (KKT) conditions for \textbf{P3} is listed as follows:
\begin{subequations}\label{eq:kkt}
\begin{numcases}{\hspace{-0.7cm}}
\beta e^{\frac{\theta}{\lambda(\gamma,\rho)}}\left(1 - \frac{\theta}{\lambda(\gamma,\rho)}\right) - 1 + \frac{\eta}{\gamma} + \eta_\tau\tau -\eta_TT = 0,\label{eq:kkt_1}\\
\eta\left(\mathbb{E}\left[\frac{\lambda(\gamma,\rho)}{\gamma}\right]- \bar{P}_\tau\right) = 0,\label{eq:kkt_2}\\
\eta_T(T\lambda(\gamma, \rho) - 1) = 0,\label{eq:kkt_3}\\
\eta_\tau(\tau\lambda(\gamma,\rho) - 1) =0,\label{eq:kkt_4}\\
\eta,\eta_\tau,\eta_T \geq 0.\label{eq:kkt_5}
\end{numcases}
\end{subequations}
We can obtain the optimal solution for \textbf{P3} via solving the above equations. The corresponding details are concluded in Theorem~\ref{thm:optimal_sampling_scheme}.

\begin{theorem}\label{thm:optimal_sampling_scheme}
The optimal sampling scheme, denoted by $\lambda_{\rm opt}(\gamma, \rho)$, for \textbf{P3} is given as follows:
\begin{equation}\label{eq:opt_sampling_scheme}
\lambda_{\rm opt}(\gamma, \rho) \!=\! \left\{ {\begin{array}{*{20}{l}}
\frac{1}{T},&\!\mbox{if~}0\leq\gamma < \gamma_1^{\rm th},\\
\frac{\theta}{1 + \mathcal{W}\left(\frac{\eta-\gamma}{e\beta\gamma}\right)},&\!\mbox{if~}\gamma_1^{\rm th}\!\leq\!\gamma\! \leq\!\gamma_2^{\rm th},\\
\frac{1}{\tau},&\!\mbox{if~}\gamma > \gamma_2^{\rm th},
\end{array}} \right.
\end{equation}
where $\gamma_1^{\rm th}$ and $\gamma_2^{\rm th}$ are two thresholds of the channel power gain equal to $\eta/(1-\beta e^{\theta T}(1-\theta T))$ and $\eta/(1-\beta e^{\theta\tau}(1-\theta\tau))$, respectively, $\mathcal{W}(\cdot)$ is the Lambert \emph{W} function~\cite{corless1996lambert}, and $\eta$ is determined by
\begin{align}\label{eq:eta_solution}
\mathbb{E}\left[\frac{\lambda_{\rm opt}(\gamma,\rho)}{\gamma}\right]=\bar{P}_\tau.
\end{align}
\end{theorem}

\begin{IEEEproof}
The slackness condition (\ref{eq:kkt_3}) indicates that, if $\eta_T > 0$, we have $\lambda(\gamma,\rho) = 1/T$. Substituting $\eta_T > 0$ and $\lambda(\gamma,\rho) = 1/T$ into (\ref{eq:kkt_1}), we have $\gamma < \eta/(1-\beta e^{\theta T}(1-\theta T))$. Similarly, based on the slackness condition (\ref{eq:kkt_4}), if $\eta_\tau > 0$, we have $\lambda(\gamma,\rho) = 1/\tau$. Substituting $\eta_\tau> 0$ and $\lambda(\gamma,\rho) = 1/\tau$ into (\ref{eq:kkt_1}), we have $\gamma > \eta/(1-\beta e^{\theta \tau}(1-\theta \tau))$. Once $\eta_\tau$ and $\eta_T$ are both equal to zero, Eq.~(\ref{eq:kkt_1}) reduces to
\begin{align}\label{thm:eq_root}
\beta e^{\frac{\theta}{\lambda(\gamma,\rho)}}\left(1 - \frac{\theta}{\lambda(\gamma,\rho)}\right) - 1 + \frac{\eta}{\gamma} = 0.
\end{align}
Solving this equation for $\lambda(\gamma,\rho)$, we have $\lambda(\gamma,\rho) = \theta /(1 + \mathcal{W}((\eta-\gamma)/(e\beta\gamma)))$, where $\mathcal{W}(\cdot)$ is the Lambert \emph{W} function~\cite{corless1996lambert}, i.e., the inverse function of $xe^x$. Finally, based on the slackness condition (\ref{eq:kkt_2}), we have Eq.~(\ref{eq:eta_solution}).
\end{IEEEproof}

In particular, when the AoI exponent approaches to zero\footnote{We will later show that when the violation probability of the peak age $\rho$ approaches one and zero, the optimal AoI exponent becomes zero and the infinite, respectively.}, $\gamma_1^{\rm th}$ and $\gamma_2^{\rm th}$ both reduce to $\eta/(1-\beta)$. Thus, the optimal sampling scheme reduces to the step function of channel power gain, i.e., we have
\begin{equation}
\lim_{\theta\rightarrow 0}\lambda_{\rm opt}(\gamma,\rho) = \left\{ {\begin{array}{*{20}{l}}
\frac{1}{T},&\mbox{if~}0\leq\gamma \leq \frac{\eta}{1-\beta},\\
\frac{1}{\tau},&\mbox{otherwise.}
\end{array}} \right.
\end{equation}
When $\theta$ goes to the infinite, we first reorganize Eq.~(\ref{thm:eq_root}) as
\begin{align}
\beta e^{\frac{\theta}{\lambda(\gamma,\rho)}}\left(1 - \frac{\theta}{\lambda(\gamma,\rho)}\right) = 1 - \frac{\eta}{\gamma},
\end{align}
and then raising to the power of $1/\theta$ on both sides as well as letting $\theta\rightarrow\infty$, we have
\begin{align}
e^{\frac{1}{\lambda(\gamma,\rho)}} = \lim_{\theta\rightarrow\infty}\eta^{\frac{1}{\theta}}.
\end{align}
Surprisingly, we find that the sampling rate is no longer affected by the channel power gain and thus it cannot be equal to $\frac{1}{T}$ or $\frac{1}{\tau}$ as the channel power gain varies. This implies that $\gamma_1^{\rm th}$ and $\gamma_2^{\rm th}$ shall converge to zero and infinite with $\theta$ close to the infinite. Therefore, the optimal sampling scheme reduces to the constant function as $\theta$ approaches to the infinite, i.e.,
\begin{equation}
\lim_{\theta\rightarrow \infty}\lambda_{\rm opt}(\gamma,\rho) = {\rm constant}.
\end{equation}

\begin{figure}
\centering
\includegraphics[scale = 0.6]{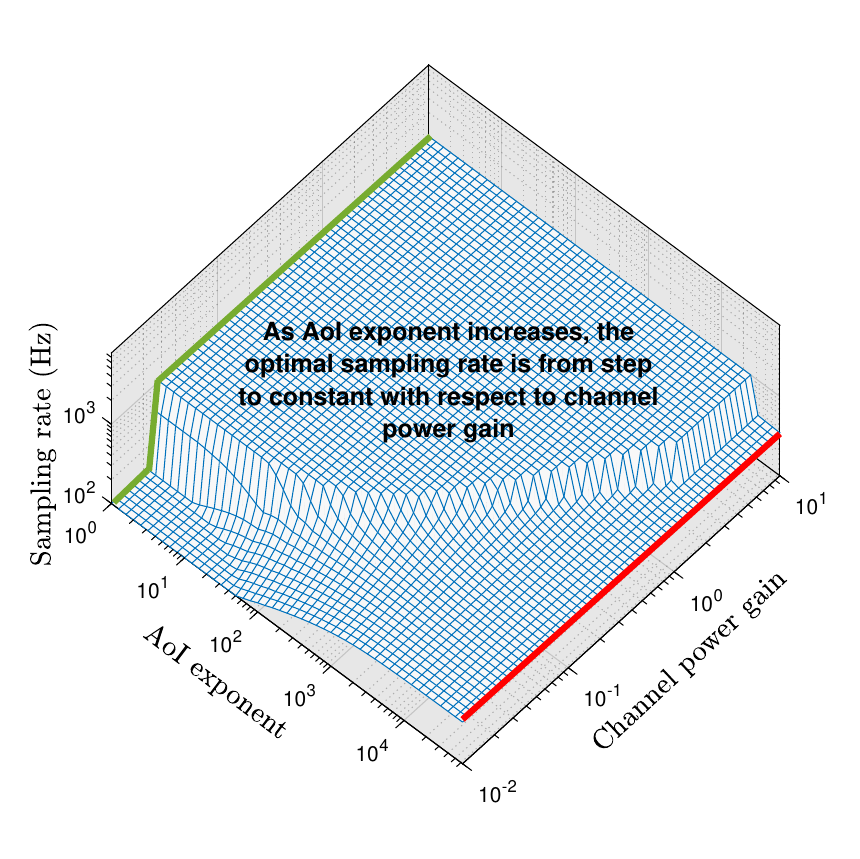}
\caption{Optimal sampling rate with respect to AoI exponent and channel power gain.}
\label{fig:optimal_sampling_rate}
\end{figure}

As mentioned before, $\lambda_{\rm opt}(\gamma, \rho)$ is also the optimal solution of \textbf{P2} once the value of $\beta$ converges. We illustrate the optimal sampling scheme in Fig.~\ref{fig:optimal_sampling_rate} to highlight its insights. It can be seen from the figure that, as the AoI exponent $\theta$ varies from zero to the infinite, the optimal sampling scheme changes from the step function of channel power gain to the constant function. Moreover, as the AoI exponent increases, $\gamma_1^{\rm th}$ decreases and $\gamma_2^{\rm th}$ increases. These phenomena comply with the above theoretical analysis.

\begin{figure}
\centering
\includegraphics[scale = 0.6]{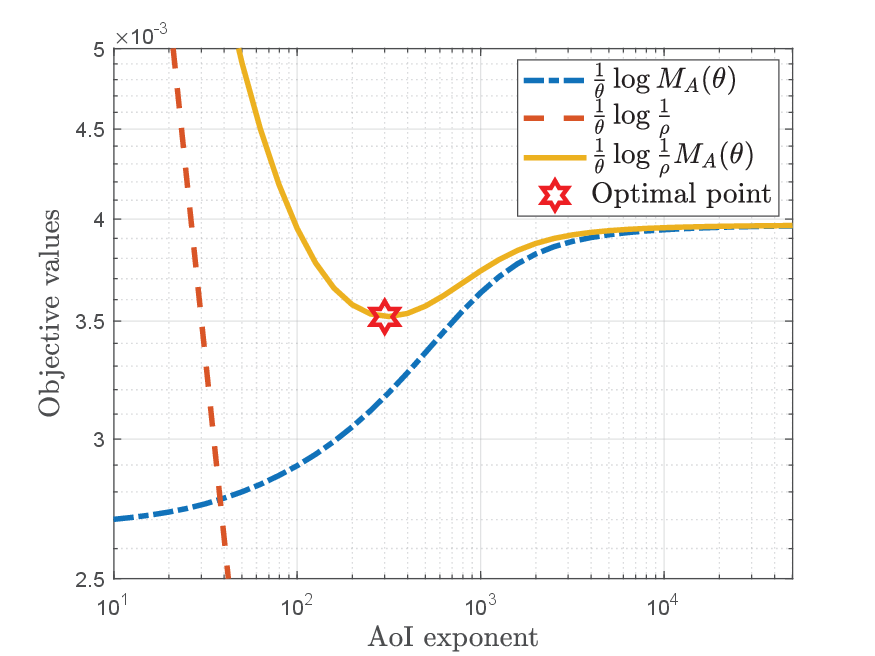}
\caption{The tendencies of $(1/\theta)\log M_A(\theta)$, $(1/\theta)\log(1/\rho)$, and $(1/\theta)\log((1/\rho)M_A(\theta))$ with respect to $\theta$.}
\label{fig:optimal_theta}
\end{figure}

Next, we discuss how to find the optimal AoI exponent with the given violation probability of the peak age. As indicated in Eq.~(\ref{eq:h_function}), $f(\theta,\lambda(\gamma,\rho))$, i.e., $1/\theta\log ((1/\rho)M_A(\theta))$, is represented as the sum of two parts, that is, $1/\theta\log M_A(\theta)$ and $1/\theta\log (1/\rho)$. To be more visualized, we plot the each part as well as the combination of them in Fig.~\ref{fig:optimal_theta}. It should be mentioned that the value of $1/\theta\log M_A(\theta)$ is calculated based on the above-derived optimal sampling scheme. We can see that $1/\theta\log M_A(\theta)$ is increasing while $1/\theta\log (1/\rho)$ is decreasing with respect to $\theta$. These two parts jointly lead to the result that $f(\theta,\lambda(\gamma,\rho))$ firstly decreases and then increases against $\theta$. This motivates us to use the bisection search method to find out the optimal AoI exponent. Suppose that the region of $\theta$ in bisection search is $[\theta_{\rm left}, \theta_{\rm right}]$. Then, we evaluate the first-order derivative of $f(\theta,\lambda(\gamma,\rho))$ at $(\theta_{\rm left} + \theta_{\rm right})/2$. If the derivative is less than zero, the value of $\theta_{\rm left}$ is updated to $(\theta_{\rm left} + \theta_{\rm right})/2$. Otherwise, we update the value of $\theta_{\rm right}$ to $(\theta_{\rm left} + \theta_{\rm right})/2$. It is worth mentioning that, when the violation probability is one, the second part of $f(\theta,\lambda(\gamma,\rho))$, that is, $1/\theta\log (1/\rho)$, reduces to zero, and thus the optimal AoI exponent is zero to minimize the first part of $f(\theta,\lambda(\gamma,\rho))$. When the violation probability is zero, the second part of $f(\theta,\lambda(\gamma,\rho))$ becomes infinite if the AoI exponent is finite, and thus the optimal AoI exponent should be infinite to eliminate its influence. With the above-derived optimal AoI exponent, the optimal sampling scheme is obtained by substituting it into Eq.~(\ref{eq:opt_sampling_scheme}).

\section{Application of Statistical AoI at Medium Access Control Layer: Multiple Critical Status Updates With Time Division Multiple Access}
\label{sec:statistical_aoi_mac}
In this section, we investigate the influence of the unreliable transmissions at the medium access control (MAC) layer on the age process. To mitigate the impact and meet diverse risk-sensitive applications' requirements, the maximum-statistical-AoI minimization problem for multi-status updates with time division multiple access~(TDMA) is studied.

\subsection{System Description}
\begin{figure}
  \centering
  \includegraphics[scale=0.86]{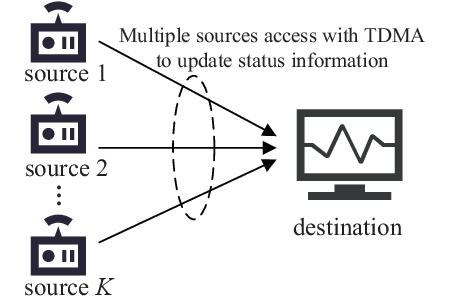}
  \caption{The system model of multi-status updates with TDMA.}\label{fig:sys_model_tdma}
\end{figure}

As shown in Fig.~\ref{fig:sys_model_tdma}, we consider a scenario where $K$ sources wish to update their respective real-time status to the target. The requirement of the peak age violation probability for source $k$ is denoted by $\rho_k$, where $k\in\mathcal{K}$ and $\mathcal{K}=\{1,2,...,K\}$. The TDMA scheme is used and the duration of a TDMA frame is $T_{\rm TDMA}$. Within each frame, the transmission time allocated to the $k$th source is denoted by $\tau_k$. For each transmission, the error transmission probability is modeled by $\varepsilon(\tau_k)$, where $\varepsilon(\tau_k)$ is a decreasing function with respect to $\tau_k$, because when less time is used, more information has to be squeezed, leading to a higher error level. Most applications for multi-status updates are low-cost sensor systems. To simplify the implementation, we assume that multiple replicas of the status packet are transmitted within each transmission time to increase transmission reliability. The error transmission probability of each replica is $p$, which is assumed to be independently identically distributed (i.i.d.) over multiple replicas. Then the error occurs when all replica transmissions of this packet fail, and thus the error transmission probability of each status packet over one frame is $p^{\frac{\tau_k}{\tau_s}}$, where $\tau_s$ is the transmission duration of one replica. For brevity, we use $e^{-c}$ to denote $p^{\frac{1}{\tau_s}}$. Then we have
\begin{align}\label{eq:error_model}
\varepsilon(\tau_k) = e^{-c\tau_k}.
\end{align}
When $\tau_s$ is sufficiently small, the transmission time $\tau_k$ can be supposed to be continuous.

\subsection{Age Process at the Destination}
\begin{figure}
\centering
\includegraphics[scale = 0.7]{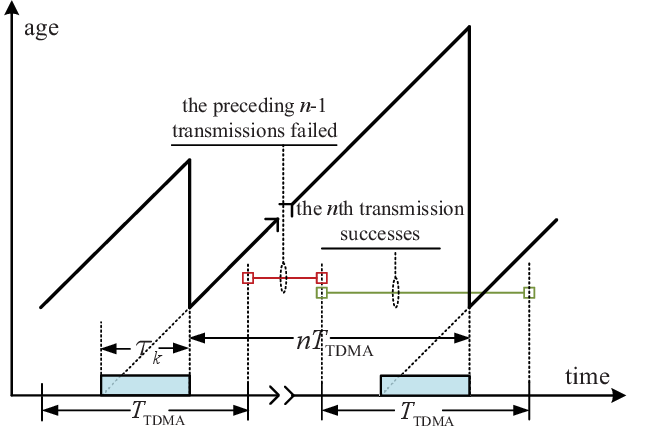}
\caption{The age process at the destination of source $k$ where $K$ sources are scheduled with TDMA.}
\label{fig:aoi_tdma}
\end{figure}
Without loss of generality, we illustrate the age process of source $k$ at the destination in Fig.~\ref{fig:aoi_tdma}. Following the definition of AoI, when the status packet is successfully received at the destination, the age is updated to $\tau_k$. Otherwise, the age continuously and linearly increases as time evolves. Then, the peak age, denoted by $A_k$, at the destination side can be given by
\begin{align}\label{eq:age_tdma}
A_k = \tau_k + nT_{\rm TDMA},~n \in \mathbb{N}_+,
\end{align}
where $n$ indicates the number of transmissions experienced until the next successful reception and $\mathbb{N}_+$ denotes the set of positive integer numbers. The probability that $A_k$ equals to $\tau_k + nT_{\rm TDMA}$ is then given by
\begin{align}\label{eq:probability_tdma}
{\rm Pr}\left(A_k = \tau_k + nT_{\rm TDMA}\right) = \varepsilon(\tau_k)^{n-1} (1 - \varepsilon(\tau_k)).
\end{align}

\subsection{Maximum-Statistical-AoI Minimization Problem}
Following the probability distribution function specified in Eq.~(\ref{eq:probability_tdma}), the MGF of $A_k$ can be presented by
\begin{align}
\mathbb{E}[e^{\theta_k A_k}] &= \sum_{n = 1}^{\infty}e^{\theta_k (\tau_k + nT_{\rm TDMA})} \varepsilon(\tau_k)^{n-1} (1 - \varepsilon(\tau_k))\\
&= \frac{(1 - \varepsilon(\tau_k))e^{\theta_k(\tau_k + T_{\rm TDMA})}}{1 - \varepsilon(\tau_k)e^{\theta_k T_{\rm TDMA}}},
\end{align}
where $\theta_k$ is the AoI exponent of source $k$ and $\varepsilon(\tau_k)e^{\theta T_{\rm TDMA}} < 1$ needs to be satisfied to assure the existence of the MGF of $A_k$. Then, based on the definition of statistical AoI specified in Eq.~(\ref{eq:definition_statistical_aoi}), the statistical AoI of source $k$ at the destination, denoted by $\Delta_k(\rho_k)$, can be presented as follows:
\begin{align}
\Delta_k(\rho_k) = \min_{\theta_k}\frac{1}{\theta_k}\log\frac{(1-\varepsilon(\tau_k))e^{\theta_k(\tau_k+T_{\rm TDMA} )}}{\rho_k(1 - \varepsilon(\tau_k)e^{\theta_k T_{\rm TDMA}})}.
\end{align}
Our objective is to obtain the optimal transmission time allocation scheme to minimize the maximum statistical AoI among $K$ sources. Towards this end, the corresponding maximum-statistical-AoI minimization problem, denoted by \textbf{P4}, is formulated as follows:
\begin{subequations}
\begin{align}
\textbf{P4}:
\min_{\tau_k} \,\,\, &\max(\Delta_1(\rho_1), \Delta_2(\rho_2), ..., \Delta_K(\rho_K)),\\
{\rm s.t.}\,\,\,
& \sum_{k = 1}^{K}\tau_k \leq T_{\rm TDMA}, \label{eq:tdma_c_1}\\
& \tau_k \geq 0,\forall k \in \mathcal{K},\label{eq:tdma_c_2}
\end{align}
\end{subequations}
where the constraint (\ref{eq:tdma_c_1}) requires that the sum of $\tau_k$ should be less than $T_{\rm TDMA}$, i.e., the total transmission time within a frame for all sources cannot exceed the frame length, and the constraint (\ref{eq:tdma_c_2}) indicates that $\tau_k$ should be greater than zero.


\begin{figure}
\centering
\subfigure[]{\includegraphics[scale = 0.6]{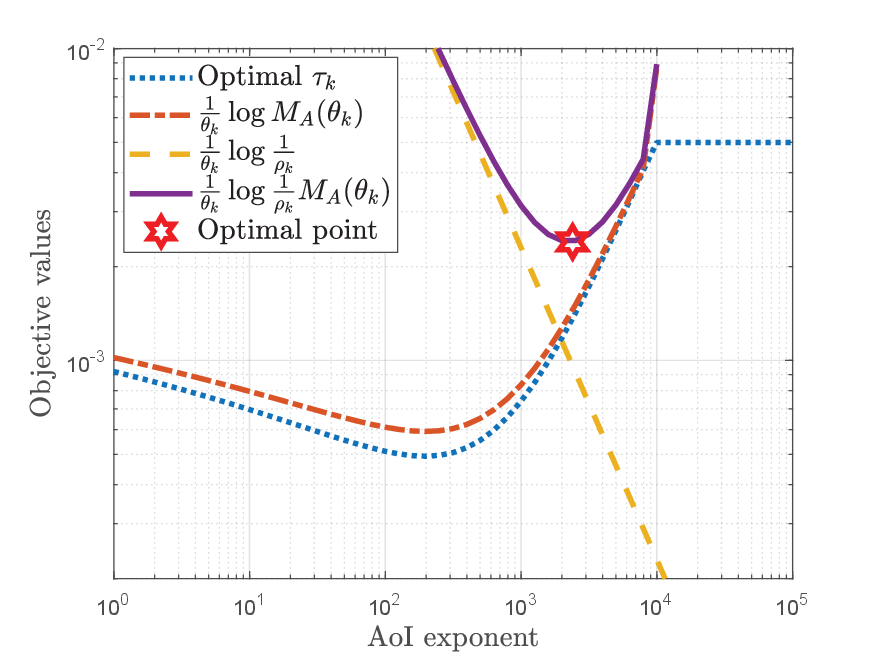}\label{fig:large_tau_max}}
\subfigure[]{\includegraphics[scale = 0.6]{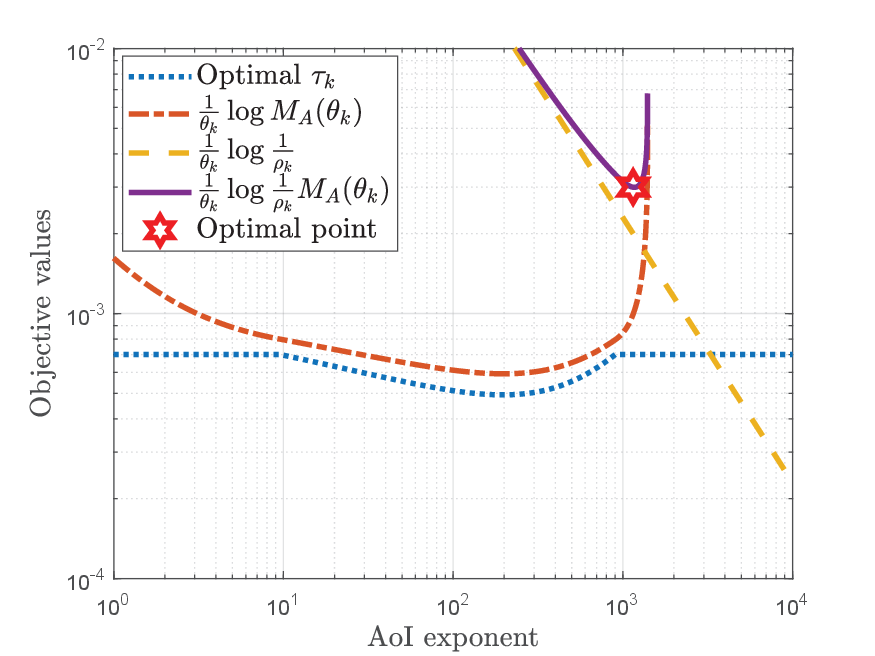}\label{fig:small_tau_max}}
\caption{The tendencies of $\tau_k$, $1/\theta_k\log M_A(\theta_k)$, $1/\theta_k\log(1/\rho_k)$, and  $(1/\theta_k)\log((1/\rho_k)M_A(\theta_k))$ with respect to $\theta_k$.}
\label{fig:optimal_theta_MAC}
\end{figure}


\subsection{Optimal Transmission Time Allocation Scheme}
It is challenging to straightforward solve \textbf{P4}, and thus we firstly consider solving one special case where $K = 1$ and then generalize the results to $K > 1$. Moreover, in practical applications, the error transmission probability is generally small, and we have $1- \varepsilon(\tau_k)\approx 1$. With the above settings, \textbf{P4} is reduced to \textbf{P5} as follows:
\begin{subequations}
\begin{align}
\textbf{P5}:
\min_{\tau_k,\theta_k} \,\,\, &\tau_k - \frac{1}{\theta_k}\log(1 - \varepsilon(\tau_k)e^{\theta_k T_{\rm TDMA}}) + \frac{1}{\theta_k}\log\frac{1}{\rho_k},\\
{\rm s.t.}\,\,\,
& 0 \leq \tau_k \leq \tau_k^{\rm max},
\end{align}
\end{subequations}
where $\tau_k^{\rm max}$ is the upper-bound of $\tau_k$ and less than $T_{\rm TDMA}$. We propose a two-step method to solve \textbf{P5}. At the first step, we regard $\theta_k$ as a constant and find the optimal $\tau_k$. Then, we search for the optimal $\theta_k$ with its associated optimal $\tau_k$. The subproblem corresponding to the first step, denoted by \textbf{P6}, is the obtained as follows:
\begin{subequations}
\begin{align}
\textbf{P6}:
\min_{\tau_k} \,\,\, &\tau_k - \frac{1}{\theta_k}\log\left(1 - \varepsilon(\tau_k)e^{\theta_k T_{\rm TDMA}}\right),\\
{\rm s.t.}\,\,\,
& 0 \leq \tau_k \leq \tau_k^{\rm max}.
\end{align}
\end{subequations}
The solution for \textbf{P6} is described in Theorem~\ref{th:opt_transmision_time}.
\begin{theorem}\label{th:opt_transmision_time}
The optimal transmission time, denoted by $\tau_k^{\rm opt}$, is determined by
\begin{align}\label{eq:opt_tau_k_final}
\tau_k^{\rm opt} = \min(\tau_k^{\rm max}, \widetilde\tau_k),
\end{align}
where
\begin{align}\label{eq:special_tau}
\widetilde\tau_k = \frac{1}{c}\log\left(1 + \frac{c}{\theta_k}\right) + \frac{\theta_k}{c}T_{\rm TDMA}.
\end{align}
\end{theorem}
\begin{IEEEproof}
For presentation convenience, we represent the objective of \textbf{P6} by $g(\tau_k)$. Taking the first-order derivative of $g(\tau_k)$ with respect to $\tau_k$, we have
\begin{align}
\frac{\partial g(\tau_k)}{\partial\tau_k} &= 1 + \frac{\varepsilon'(\tau_k)e^{\theta_k T_{\rm TDMA}}}{\theta_k(1 - \varepsilon(\tau_k)e^{\theta_k T_{\rm TDMA}})},
\end{align}
where $\varepsilon'(\tau_k)$ is the first-order derivative of $\varepsilon(\tau_k)$. By setting $\partial g(\tau_k)/\partial\tau_k = 0$, we have
\begin{align}\label{eq:opt_tau}
\theta_k\varepsilon(\tau_k)-\varepsilon'(\tau_k) = \theta_k e^{-\theta_k T_{\rm TDMA}}.
\end{align}
Then, substituting $\varepsilon(\tau_k) = e^{-c\tau_k}$ given by Eq.~(\ref{eq:error_model}) into the above equation, we obtain Eq.~(\ref{eq:special_tau}). Since $\tau_k \leq \tau_k^{\rm max}$, the optimal solution for \textbf{P6} is $\min(\tau_k^{\rm max}, \widetilde\tau_k)$.
\end{IEEEproof}

To facilitate understanding of Theorem~\ref{th:opt_transmision_time}, we draw typical cases of the optimal transmission time $\tau_k$ with respect to AoI exponent in Fig.~\ref{fig:optimal_theta_MAC}. As shown in Fig. \ref{fig:large_tau_max}, the optimal $\tau_k$ firstly decreases, then increases, and finally converges to a constant as the AoI exponent increases. With the optimal $\tau_k$, we also illustrate the tendencies of $1/\theta_k\log M_A(\theta_k)$, $1/\theta_k\log(1/\rho_k)$, and $1/\theta_k\log((1/\rho_k)M_A(\theta_k))$ with respect to $\theta_k$. It can be seen from the figure that, there exists the optimal AoI exponent, which can be obtained by substituting Eq.~(\ref{eq:special_tau}) into $1/\theta_k\log((1/\rho_k)M_A(\theta_k))$ and then taking the derivative of it with respect to $\theta_k$ with setting the result to be zero, i.e., we have
\begin{align}\label{eq:optimal_theta_k}
\frac{1}{c}T_{\rm TDMA} - \frac{1}{\theta_k^2}\log\frac{1}{\rho_k} - \frac{1}{\theta_k^2}\log\frac{c+\theta_k}{c} = 0.
\end{align}
Because $1/\theta_k^2\log((c+\theta_k)/c)$ is very close to zero\footnote{When $\theta_k$ is small, $\log((c+\theta_k)/c)$ is close to zero. When $\theta_k$ is large, $1/\theta_k^2$ is close to zero.}, the corresponding influence on Eq.~(\ref{eq:optimal_theta_k}) can be ignored. Then, the optimal AoI exponent, denoted by $\theta_k^{\rm opt}$, is approximated as
\begin{align}\label{eq:opt_theta_large_tau}
\theta_k^{\rm opt} \approx \sqrt{\frac{c\log\frac{1}{\rho_k}}{T_{\rm TDMA}}}.
\end{align}
Plugging $\theta_k^{\rm opt}$ into Eq.~(\ref{eq:special_tau}), we update $\widetilde\tau_k$ to
\begin{align}\label{eq:opt_tau_k}
\widetilde\tau_k \approx \frac{1}{c}\log\left(1 + \sqrt{\frac{cT_{\rm TDMA}}{\log\frac{1}{\rho_k}}}\right) + \sqrt{\frac{T_{\rm TDMA}\log\frac{1}{\rho_k}}{c}}.
\end{align}

It should be mentioned that the above results are based on the premise of $\widetilde\tau_k \leq \tau_k^{\rm max}$. In the following, we discuss the case of $\widetilde\tau_k > \tau_k^{\rm max}$. Fig.~\ref{fig:small_tau_max} shows the tendency of $1/\theta_k\log((1/\rho_k)M_A(\theta_k))$ with respect to $\theta_k$ for $\widetilde\tau_k > \tau_k^{\rm max}$, where the minimum statistical AoI is achieved when $\tau_k = \tau_k^{\rm max}$. Substituting $\tau_k = \tau_k^{\rm max}$ into the objective of \textbf{P5}, denoted by $h(\theta_k,\tau_k^{\rm max})$, and taking the derivative of $h(\theta_k,\tau_k^{\rm max})$ with respect to the AoI exponent, we have
\begin{multline}
\frac{\partial h(\theta_k,\tau_k^{\rm max})}{\partial \theta_k} = \frac{1}{\theta_k^2}\log(1 - e^{-c\tau_k^{\rm max}+ \theta_k T_{\rm TDMA}})\\
+\frac{T_{\rm TDMA}e^{-c\tau_k^{\rm max}+\theta_k T_{\rm TDMA}}}{\theta_k(1 - e^{-c\tau_k^{\rm max}+\theta_k T_{\rm TDMA}})}-\frac{1}{\theta_k^2}\log\frac{1}{\rho_k}.
\end{multline}
The closed-form solution of $\theta_k$ to the equation $\partial h(\theta_k,\tau_k^{\rm max})/\partial \theta_k = 0$ is intractable. Alternatively, we can use the bisection search method to quickly obtain the numerical solution.

So far, we have shown the way to identify the optimal transmission time as well as AoI exponent for the case of $K = 1$, where the optimal transmission time is not always $\tau_k^{\rm max}$. The reason is that although the longer transmission time can enhance the reliability of status information, it also increases the age considerably. The transmission time specified in Eq.~(\ref{eq:opt_tau_k_final}) can achieve the balance between reliability and age via minimizing the statistical AoI.

\begin{figure}
\centering
\includegraphics[scale = 0.6]{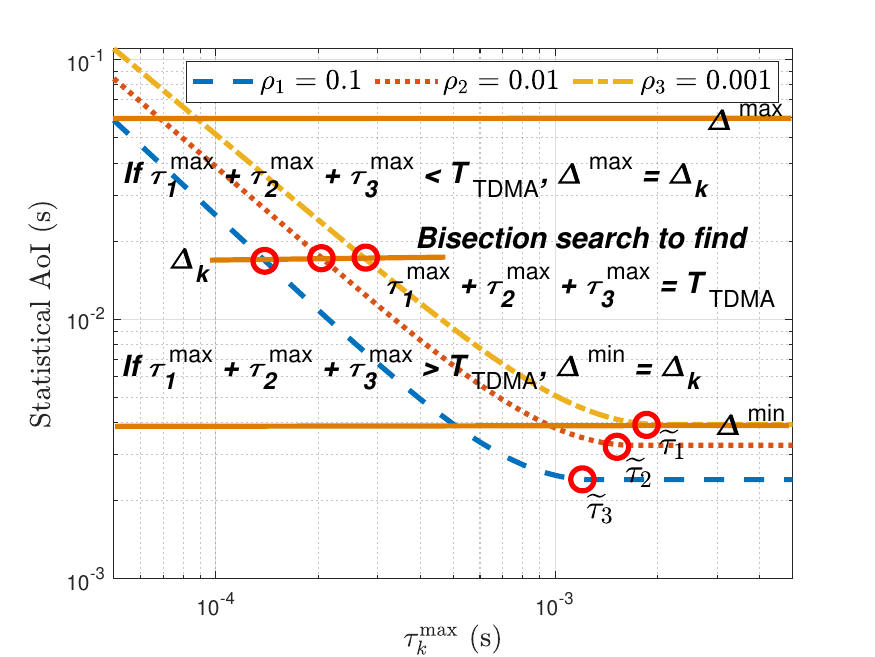}
\caption{The statistical AoI versus $\tau_k^{\rm max}$ with different violation probabilities of the peak age.}
\label{fig:statistical_AoI_tau_max}
\end{figure}

Depending on the above theoretical results, we can readily calculate the value of statistical AoI with respect to $\tau_k^{\rm max}$. As illustrated in Fig.~\ref{fig:statistical_AoI_tau_max}, with the given violation probability of the peak age, the statistical AoI firstly decreases and then keeps constant as $\tau_k^{\rm max}$ increases from zero to $T_{\rm TDMA}$, where the turning point is $\widetilde\tau_k$. Also, it can be seen that the smaller violation probability corresponds to the larger statistical AoI. Following this thought, we share an effective way to resolve \textbf{P4}. Because \textbf{P4} is a min-max problem, if $\Delta_k(\rho_k) < \Delta_j(\rho_j)$, $k,j\in \mathcal{K}$, the originally allocated transmission time for source $k$ can be partly assigned to source $j$. Then, $\Delta_k(\rho_k)$ increases and $\Delta_j(\rho_j)$ decreases, leading to that the objective of \textbf{P4} further decreases. Therefore, in most cases, the solution to \textbf{P4} has the following relations hold:
\begin{align}
\Delta_1(\rho_1) =  \Delta_2(\rho_2) = ... = \Delta_K(\rho_K).
\end{align}
As such, we can use the bisection search method to find out the minimum maximum-statistical-AoI as well as the corresponding $(\tau_1, \tau_2, ..., \tau_K)$. Specifically, as shown in Fig.~\ref{fig:statistical_AoI_tau_max}, we take an example that $K = 3$ and the corresponding $\rho_1$, $\rho_2$, and $\rho_3$ are set to 0.1, 0.01, 0.001, respectively. The region of $\Delta_k$ in bisection search is $[\Delta^{\min}, \Delta^{\max}]$. Then, the value of $\tau_k^{\rm max}$ associated with $\Delta_k = (\Delta^{\min} + \Delta^{\max})/2$ is calculated for $k$ from $1$ to $3$. Next, if $\sum_{k = 1}^{K} \tau_k^{\rm max} > T_{\rm TDMA}$, $\Delta^{\min}$ is updated to $\Delta_k$. Otherwise, we update $\Delta^{\max}$ to $\Delta_k$. This procedure is repeated iteratively until $\sum_{k = 1}^{K} \tau_k^{\rm max}$ converges to $T_{\rm TDMA}$.

\section{Numerical and Simulation Results}
\label{sec:numerical_results}
In this section, we evaluate performances of our proposals using the statistical AoI as the objective to optimize the status update systems. The numerical results consist of two subsections: a) the sampling optimization scheme at the physical layer and b) the transmission time optimization scheme at the MAC layer, respectively. The default settings of the simulation parameters are listed in Table~\ref{tab:table_2}.
\begin{table}
\caption{Values of Simulation Parameters}
\small
\label{tab:table_2}
\centering
\begin{tabular}{|c|c|}
\hline
\textbf{Parameter} & \textbf{Value}\\
\hline
\hline
average transmit power $\bar{P}$ & 0.1 W\\
\hline
bandwidth $B$ & 1 MHz\\
\hline
size of status packet $D$ & 100 bits\\
\hline
distribution of channel power gain $\gamma$ & Rayleigh distribution\\
\hline
average channel power gain $\bar{\gamma}$ & 1\\
\hline
transmission time of one status packet $\tau$ & 1 ms\\
\hline
channel coherence time $T$ & 0.1 s\\
\hline
number of sources $K$ & 3 \\
\hline
error factor $c$ & 1000 \\
\hline
duration of one TDMA frame $T_{\rm TDMA}$ & 10 ms\\
\hline
\end{tabular}
\end{table}

\subsection{Sampling Optimization Scheme at The Physical Layer}

\begin{figure}
\centering
\includegraphics[scale = 0.6]{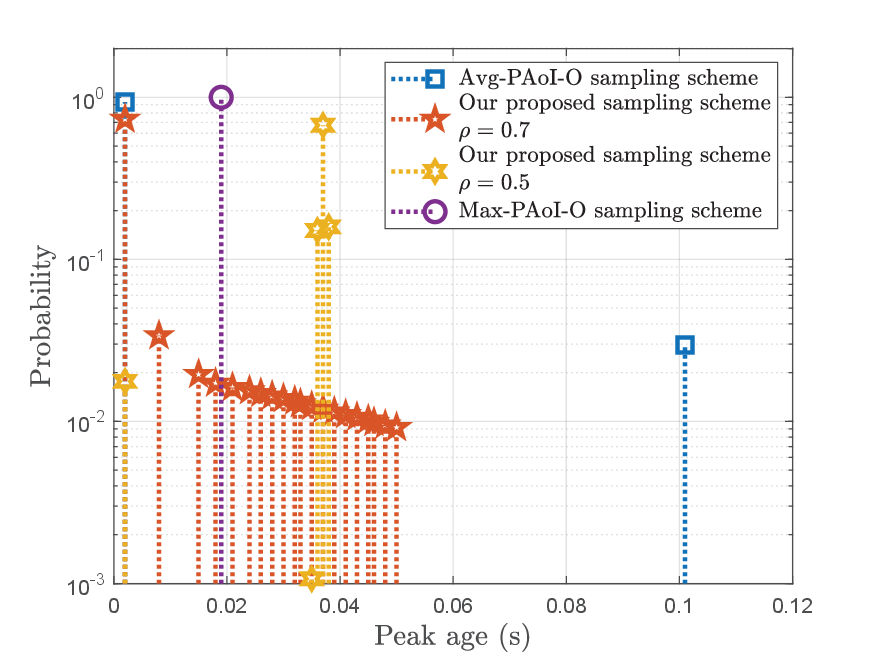}
\caption{The probability distribution of the peak age under Avg-PAoI-O sampling scheme, Max-PAoI-O sampling scheme, and our proposed sampling scheme with different risk-sensitive requirements.}
\label{fig:pdf_PHY_scheme}
\end{figure}

\begin{figure}
\centering
\includegraphics[scale = 0.6]{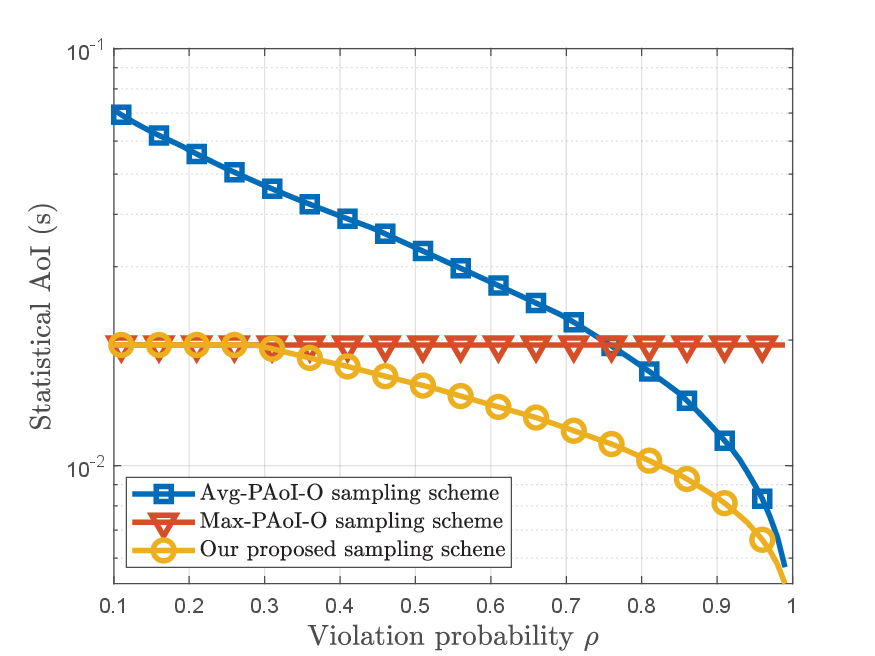}
\caption{The performance comparison among our proposed sampling scheme, Avg-PAoI-O sampling scheme, and Max-PAoI-O sampling scheme.}
\label{fig:performance_statistical_AoI_PHY}
\end{figure}

Figure~\ref{fig:pdf_PHY_scheme} shows the probability distribution of the peak age under the average-peak-AoI-oriented (Avg-PAoI-O) sampling scheme, the maximum-peak-AoI-oriented (Max-PAoI-O) sampling scheme, and our proposed sampling scheme with different risk-sensitive requirements. Avg-PAoI-O sampling scheme and Max-PAoI-O sampling scheme stand for the schemes derived by using the average peak AoI and the maximum peak AoI as the objective, respectively. It can be seen form the figure that the overlarge peak age exists when using Avg-PAoI-O sampling scheme because the metric average peak AoI is risk-insensitive. When using Max-PAoI-O sampling scheme, the peak age is a constant around 0.02 s. With using our proposed sampling scheme by setting the violation probability $\rho$ as 0.7, the peak age is distributed around 0.002 s with a high probability. Meanwhile, the maximum peak age is not very large. When the violation probability $\rho$ is set to 0.5, the probability distribution of the peak age is further concentrated.

Figure~\ref{fig:performance_statistical_AoI_PHY} shows the performance of our proposed sampling scheme, Avg-PAoI-O sampling scheme, and Max-PAoI-O sampling scheme, respectively. It can be seen that as the violation probability increases, the statistical AoI of Avg-PAoI-O sampling scheme decreases. In contrast, the statistical AoI with Max-PAoI-O sampling scheme stays constant regardless of variation of the violation probability. The reason behind this observation is that the sampling rate in Max-PAoI-O sampling scheme is a constant. In comparison, our proposed scheme can always achieve the minimum statistical AoI, which firstly keeps constant and then begins to decrease against the violation probability.

\subsection{Transmission Time Optimization Scheme at The MAC Layer}

\begin{figure}
\centering
\includegraphics[scale = 0.6]{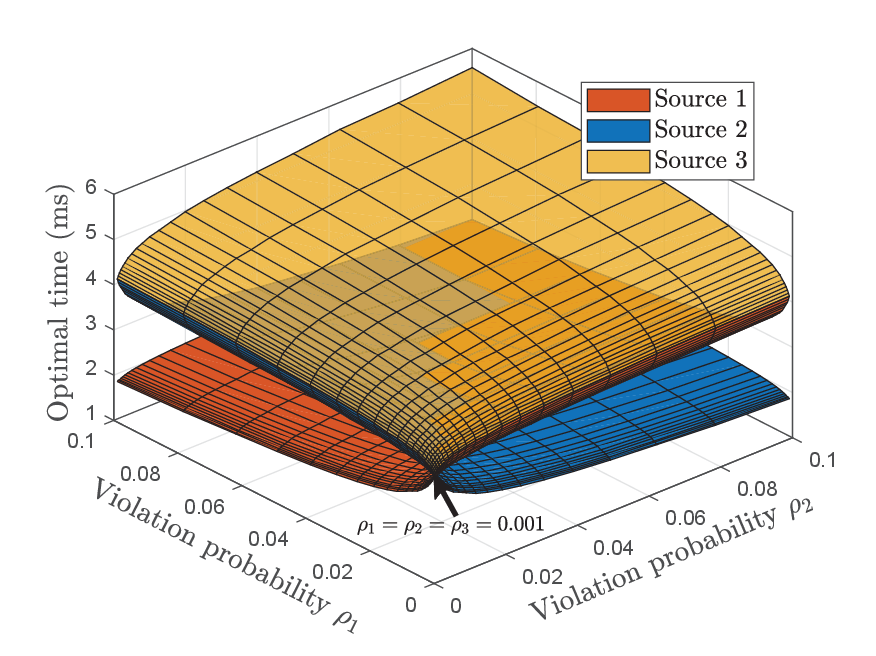}
\caption{Optimal transmission time for each source with their respective violation probability requirements.}
\label{fig:scheme_MAC}
\end{figure}

\begin{figure}
\centering
\includegraphics[scale = 0.6]{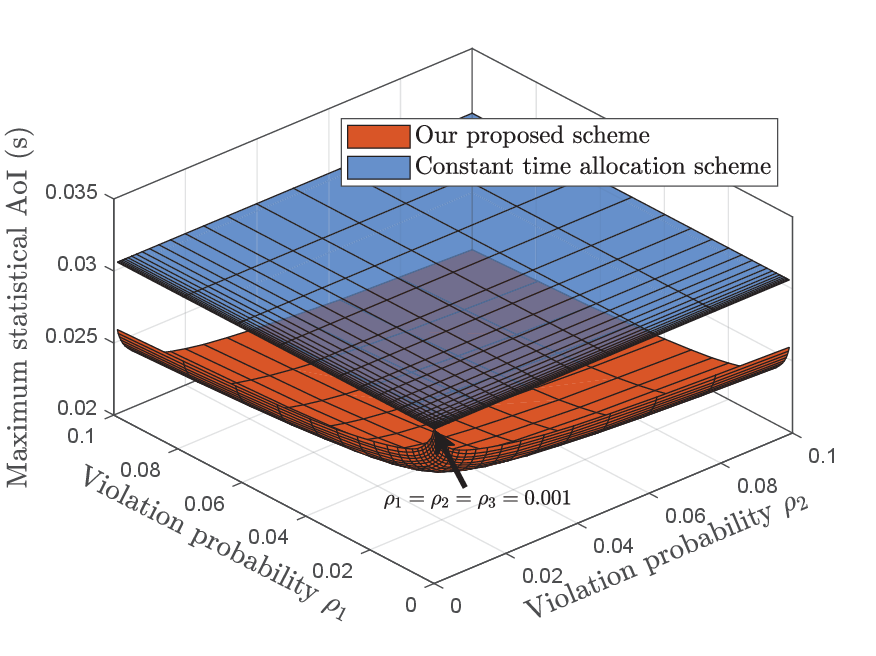}
\caption{The maximum statistical AoI with respect to various violation probability requirements of the sources under our proposed scheme and the constant time allocation scheme.}
\label{fig:performance_MAC}
\end{figure}

Figure~\ref{fig:scheme_MAC} illustrates the optimal transmission time of each source with their respective violation probability requirements. The violation probabilities of source 1 and source 2 vary from 0.001 to 0.1, and the violation probability of source 3 is fixed at $\rho_3=0.001$. It shows that as the violation probability of user 1 and/or user 2 decreases, the optimized transmission time of source 1 and/or source 2 increases accordingly. In the meantime, the transmission time of source 3 decreases. When $\rho_1=\rho_2=\rho_3=0.001$, the transmission time for each user becomes the same. The observation implies that more transmission time should be allocated to the source whose violation probability requirement is stricter.

Figure~\ref{fig:performance_MAC} depicts the maximum statistical AoI among three sources with respect to different violation probability requirements under our proposed scheme and the constant time allocation scheme, which means the equal transmission time allocated to each source. The violation probability of source 3 is set to $\rho_3=0.001$. The violation probabilities of sources 1 and 2 vary from 0.001 to 0.1. As shown in the figure, for our proposed scheme, the corresponding maximum statistical AoI increases as the violation probability of source 1 and/or source 2 decreases. In contrast, the maximum statistical AoI corresponding to the constant time allocation scheme keeps as a fixed value regardless of the variation of the violation probabilities of source 1 and/or source 2. The reason is that the statistical AoI of source 3 whose violation probability is the smallest dominates the maximum statistical AoI. Only when $\rho_1=\rho_2=\rho_3=0.001$, the performance of our proposed scheme is the same as that of the constant time allocation scheme.

\begin{figure}
\centering
\includegraphics[scale = 0.6]{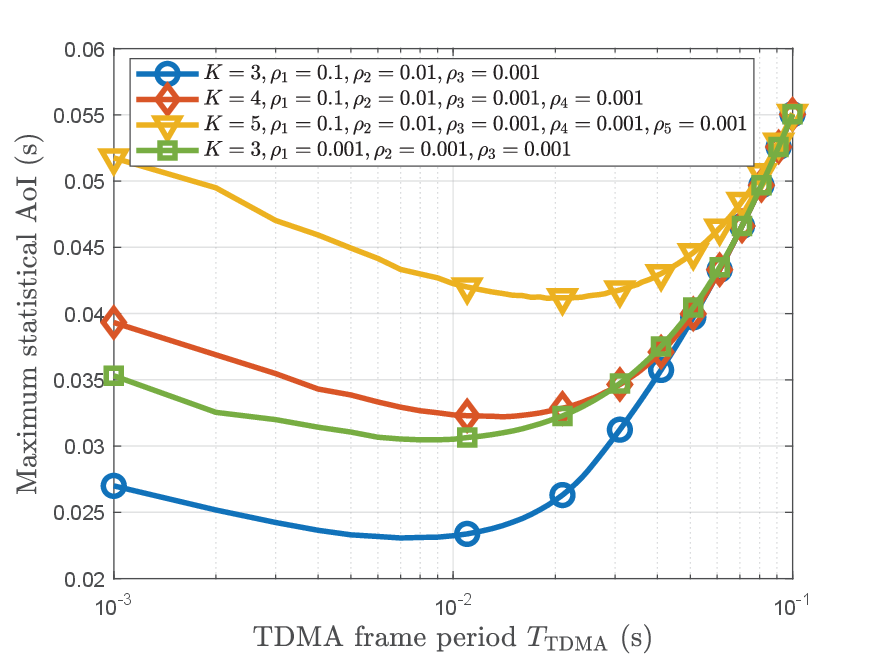}
\caption{The maximum statistical AoI with respect to TDMA frame period under different numbers of sources and violation probabilities.}
\label{fig:opt_Ttdma_MAC}
\end{figure}

Figure~\ref{fig:opt_Ttdma_MAC} illustrates the maximum statistical AoI with respect to TDMA frame period under different numbers of sources as well as violation probabilities. It is shown that as duration of TDMA frame increases, the maximum statistical AoI firstly decreases and then increases. This observation implies that there exists the optimal TDMA frame period for specific scenarios. The reason is that, on the one hand, smaller TDMA frame period can degrade transmission reliability and then more transmissions is needed for one successful status update. On the other hand, the larger TDMA frame period directly enlarges the age. The optimal TDMA frame period is affected by number of sources. We can see from the figure that as number of sources increases, the optimal TDMA frame period increases.

\section{Conclusions}
\label{sec:conclusions}
In this paper, we introduced the concept of statistical AoI, which provides a unified framework to guide the design and evaluate the information freshness performance for various risk-sensitive status update systems. To demonstrate the insights of this new risk-aware metric, the statistical AoI minimization problems for status update over wireless fading channels were investigated, and the corresponding closed-form solution was obtained. The results showed that as the requirement on the violation probability of the peak age varies from loose to strict, the optimal sampling scheme varies from the step function to the constant function of the channel power gain. How we can use statistical AoI for multiple risk-sensitive status updates was also presented. The numerical results confirm that our proposals can better meet the requirements of risk-sensitive status updates and have great potential to improve the corresponding information freshness compared to existing approaches.

\bibliographystyle{IEEEtran}
\bibliography{References}

\end{document}